# Facilitating the Integration of Ethical Reasoning into Quantitative Courses: Stakeholder Analysis, Ethical Practice Standards, and Case Studies

Rochelle E. Tractenberg[1], Suzanne Thornton[2]
[1]Collaborative for Research on Outcomes and –Metrics; and Departments of Neurology; Biostatistics, Bioinformatics & Biomathematics; and Rehabilitation Medicine, Georgetown University, Washington, DC, USA
[2]Department of Mathematics and Statistics, Swarthmore College
Swarthmore, PA, USA

**Abstract**
Case studies are typically used to teach "ethics", but emphasize narrative. When the content of a course is focused on formulae and proofs, it can seem divergent or distracting – for both instructor and learner - to introduce a case analysis. Moreover, case analyses are typically focused on issues relating to people: obtaining consent, dealing with research team members, and/or potential institutional policy violations. While relevant to some research, not all students in quantitative courses plan to become researchers, and ethical practice – of mathematics, statistics, data science, and computing – is an essential topic regardless of whether or not the learner intends to do research. It is a mistake to treat "training in ethical practice" and "training in responsible conduct of research" as the same thing, just as incorrect as it is to assume that "training in ethical practice" is the same irrespective of what the learner will actually be practicing. Ethical reasoning is a way of thinking that requires the individual to assess what they know about a potential ethical problem – their prerequisite knowledge, and in some cases, how behaviors they observe, are directed to perform, or have performed, diverge from what they know to be ethical behavior. Ethical reasoning is a learnable, improvable set of knowledge, skills, and abilities that enable learners to recognize what they do and do not know about what constitutes "ethical practice" of a discipline, and in some cases, to contemplate alternative decisions about how to first recognize, and then proceed past, or respond to, such divergences. A stakeholder analysis is part of prerequisite knowledge, and can be used whether there is or is not an actual case or behavior/situation to react to. When teaching courses with primarily quantitative content, a stakeholder analysis is a useful tool for instruction and assessment of learning. It can be used to both integrate authentic ethical content and encourage careful quantitative thought. This paper discusses how to introduce ethical reasoning, stakeholder analysis, and ethical practice standards authentically in quantitative courses.

## 1. Introduction

"**Ethics** refers to standards and practices that tell us how human beings ought to act in the many situations in which they find themselves—as friends, parents, children, citizens, businesspeople, professionals, and so on. Ethics is also concerned with our character. It requires knowledge, skills, and habits." This definition of ethics comes from the Markkula Center for Applied ethics at Santa Clara University, where a wide variety of tools and resources for promoting ethics (and applied ethics) are created and housed. However, "teaching ethics" is too vague of a task for any individual instructor teaching statistics or data science in higher education. Recently, encouragement to incorporate "ethics" into the curricula for statistics (ASA, 2014) and for data science (National Academies, 2018) has been published for undergraduate degree programs. The ASA *Curriculum Guidelines for Undergraduate Programs in Statistical Science* state that, "ethical issues should be incorporated throughout a program" (p. 13), i.e., the program remains focused on statistical science and "ethical issues" should be encountered "throughout" that program. The Guidelines for Assessment and Instruction in Statistics Education (GAISE) in Statistics Education (GAISE) College Report (GAISE College Report Revision Committee, 2016) also includes a goal, among the "desired result of all introductory statistics courses", that specifically targets "an awareness of *ethical issues* associated with sound statistical practice" (emphasis in original; p. 8). In a nuanced difference, the National Academies of Sciences, Engineering, and Medicine (National Academies 2018) seems to go further by integrating ethics into the definition of data science: "data science spans a broad(er) array of activities that involve applying principles for data collection, storage, integration, analysis, inference, communication, *and ethics*". (emphasis added, p. 1) This inclusion of ethics into the definition of the domain is intentional; the authors of this report, the Committee on Envisioning the Data Science Discipline, note that their emphasis (on ethics in the definition of data science):

"… *underscores the centrality of studying the many ethical considerations that arise as workers engage in data science*. These considerations include deciding what data to collect, obtaining permissions to use data, crediting the sources of data properly, validating the data's accuracy, taking steps to minimize bias, safeguarding the privacy of individuals referenced in the data, and using the data correctly and without alteration. It is important that students learn to recognize ethical issues and to apply a high ethical standard." (emphasis added, p. 2).

Two of the 2018 National Academies report action items (p.3) follow directly from the Committee's focus on the centrality of ethical professional practice in data science:

*Recommendation 2.4:* Ethics is a topic that, given the nature of data science, students should learn and practice throughout their education. Academic institutions should ensure that ethics is woven into the data science curriculum from the beginning and throughout.

*Recommendation 2.5:* The data science community should adopt a code of ethics; such a code should be affirmed by members of professional societies, included in professional development programs and curricula, and conveyed through educational programs. The code should be reevaluated often in light of new developments.

As has been argued elsewhere (Tractenberg 2022-A, 2022-B; Tractenberg, 2020; Hogan et al. 2017), the ASA Ethical Guidelines for Statistical Practice (ASA 2022) represent ethical practice standards that comprise both statistics and data science. In particular, the concordance of the professional practitioner communities for statistics (i.e., the ASA) and computing (Association of Computing Machinery, ACM, 2018) have longstanding ethical practice standards that are highly aligned (2018 versions discussed in Tractenberg 2020; 2018 ACM and 2022 ASA versions discussed in Tractenberg 2022-A, 2022-B). To the extent that the ASA and ACM describe the ethical practice standards of professionals in computing, statistics, and data science, "the data science community" already has access to the ASA Guidelines to utilize as its "code of ethics". Clearly, the professional and scientific communities are placing increasing value and importance on the integration of "ethics" or ethical content into the curriculum for degrees in statistics and data science. This content exists. A remaining challenge is how to integrate both the content and how to reason with it into the entire curriculum (in statistics and/or data science). Whenever courses in this curriculum are primarily quantitative, featuring formulae, theories, and proofs rather than applications, the opportunities to integrate ethical practice content appear to be limited.

## 2. Options for "teaching ethics"

The ASA (2014), GAISE (2016), and National Academies (2018) recommendations are not specific about what, or how, ethical content or "ethical issues" can effectively be integrated into either a single course (e.g., a target for an introductory statistics course; GAISE 2016) or the curriculum (ASA 2014; National Academies 2018). As has been discussed previously (Tractenberg, 2016?), simply attaching the ethical practice standards (ASA, ACM) to the syllabus will not be effective. Limiting instruction to in-class discussion of examples of unethical application(s) of statistics (e.g., unsubstantiable analyses supporting claims of voter fraud in 2020 election; deploying algorithms to steal Facebook data from unwilling or unknowing data donors in 2016), and how they violate laws, rules, or ethical practice standards definitely engages student attention. These examples also highlight the importance of learning statistics and data science - so, they don't simply present a social or cultural "good" that is both abstract and possibly unrelated to quantitative content (e.g., beneficence and nonmaleficence; Beauchamp & Childress 1984). In order to inculcate students with a sense of responsibility for ethical statistical and data science practice, a coherent approach to teaching, and to assessing learning, is needed throughout a semester or term.

### 2.1 Case studies

Case studies are typically used – and recommended - to teach "ethics". Cases that are available (the development and sharing of many cases have been supported by the National Science Foundation and National Institutes of Health) are universally contextualized well beyond formulae and proofs of theorems, lemmas, or corollaries. The National Institutes of Health Office of Intramural Research describes case studies for teaching/learning ethics this way: "*Research Ethics Cases are a tool for discussing* <u>*scientific integrity*</u>. *Cases are designed to confront the readers with a specific problem that does not lend itself to easy answers. By providing a focus for discussion, cases help staff involved in research to define or refine their own standards, to appreciate alternative approaches to identifying and resolving ethical problems, and to develop skills for dealing with hard problems on their own*."[1] (emphasis added.)

---

[1] https://oir.nih.gov/sourcebook/ethical-conduct/responsible-conduct-research-training/annual-review-ethics-case-studies

In addition to the fact that all those who teach statistical and data science practices have an obligation to teach ethical statistical and data science practice (i.e., the topic of ASA Principle Elements A12, G, and G1; ASA, 2022), the use of case studies as defined by the NIH can be generally challenging across statistical instruction for several reasons:

1. Unless the instructor is willing to add instruction/dedicate time to providing feedback on narrative responses from students (e.g., essays, open-ended responses), there is no way to effectively assess if/how students are learning or progressing.
2. Without written work reflecting the students' engagement with the case, the instructor has no way to assess learning.
3. Cases may present interesting ethical cases *without* relating in an authentic way to applying the ethical content.
4. If a course does not utilize narrative – in content or in assessment – then teaching and assessing learning with case studies might be very different from what instructors and students are familiar with. This can create challenges for students as well as for the instructors.

Moreover, it is essential to point out that the National Academies and ASA recommendations are <u>not</u> about <u>research</u> ethics but about *ethical practice*. This means that materials currently available to teach "research ethics" – none of which have been created to support quantitative instructors or courses with primarily quantitative content – will be useful for teaching "ethical practice". That is, special consideration for how to teach practitioners to be ethical when they use mathematics or statistics is fundamentally different from training researchers in "responsible conduct of research".

When the content of a course is focused on formulae and proofs, it can seem divergent or distracting to introduce a case analysis. Moreover, case analyses are typically focused on issues relating to people: obtaining consent, dealing with research team members, and/or potential institutional policy violations. Finally, cases describe actions taken by multiple parties with different roles. This is certainly one way to help students learn about ethical guidelines (or applicable rules, policies, and laws) – as they determine which of these were violated (or might have been threatened) by actors in the case. However, this approach has no way to help students appreciate how ethical practice standards -such as the ASA Ethical Guidelines for Statistical Practice (or other rules) can be used to practice ethically in general. The case approach emphasizes identification of "what went wrong" -i.e., require the learner to understand the context of scientific application of what they are in the course to begin learning, and then typically, the articulation of a response. The structure of cases (as defined by the National Institutes of Health) is to be intentionally ambiguous and not include objectively right or wrong responses. Thus, "what to do" when faced with similar situations may not be made clear to learners, or, multiple options may be presented and a clear way to make a decision about what to do in response might again be unclear.

In addition to the fact that beginners or those who are starting to integrate their quantitative knowledge might not appreciate the multitude of complexities involved in participating in a scientific project or team, teaching "responsible conduct of research" with cases that are ill-defined makes it difficult to consider how to identify similar - also ill-defined - situations in the future. Moreover, cases that present completed events and ask students to determine "what went wrong" helps to promote the idea that avoiding bad behavior, or avoiding being caught in bad behavior, is what constitutes ethical research. A final problem with typical case-based teaching of ethics is that the majority of materials that exist were developed for academic and research situations specifically. Newer resources for "ethical AI", such as those from the Princeton Dialogues on AI and Ethics (https://aiethics.princeton.edu/case-studies/case-study-pdfs/), are not specific to research, but they do feature "ethical issues" that are general. Unfortunately, like older cases, these do not relate to "ethical practice". They also tend to be self-contained, and unrelated/not relatable to quantitative material.

While quantitative courses are undoubtedly relevant to students in research or science majors, not all students in quantitative courses plan to become researchers, and ethical practice – of mathematics, statistics, data science, and computing – is an essential topic regardless of whether or not the learner intends to do research. In addition to the four classroom-based, instructor-specific difficulties that arise when quantitative course instructors want to integrate ethical content into their course, the available cases will often be quite far from being both useful and accessible to these instructors.

### 2.2 Teaching Ethical Reasoning instead of "ethics"

> "**Ethics is the effort to guide one's conduct with careful reasoning**. One cannot simply claim "X is wrong."; Rather, one needs to claim "X is wrong because (fill in the blank)"." (Briggle & Mitcham, 2012, p. 38). (emphasis added)

*Ethical reasoning* is a process that can be learned and improved. It can be taught *and assessed.* There are six elements of knowledge, skills, and abilities (KSAs) that are required for ethical reasoning (Tractenberg & FitzGerald, 2012; Tractenberg 2022-A):

1. Identify and 'quantify' prerequisite knowledge
2. Identify decision-making frameworks.
3. Identify or recognize the ethical issue.
4. Identify and evaluate alternative actions (on the ethical issue).
5. Make and justify a decision.
6. Reflect on the decision.

While it has been argued that ethical reasoning (rather than ethics) is what statistics and data science curricula should be teaching and emphasizing, it can be challenging to get to, or beyond, KSA #3 (identify or recognize the ethical issue) when the course is featuring formulae, proofs, and theorems. Critically, *proofs and formulae typically stipulate assumptions and approximations. Whenever these assumptions and approximations do not hold, cannot be met, or cannot be verified, it creates an opportunity to consider what the impacts of these failures might be*. Even in a superficially abstract and quantitative context like a proof, the importance of assumptions and approximations can be leveraged to introduce ethical practice content.

As noted, **ethical reasoning** is a way of thinking that requires the individual to assess what they know about a potential ethical problem – their prerequisite knowledge, and in some cases to contemplate alternative decisions for responding to that problem. However, when learning quantitative content, reasoning, and skills (e.g., in mathematical statistics, mathematics, calculus, etc.), it can be distracting to students as well as instructors to have to divert attention to person-based considerations like behavior. There are six elements of knowledge, skills, and abilities (KSAs) that are required for ethical reasoning. These are described in the context of mathematical content (i.e., formulae, proofs, theorems):

1. **Identify and 'quantify' prerequisite knowledge**: a practitioner should have sufficient familiarity with the ASA Ethical Guidelines for Statistical Practice (GLs) to identify at least one of the 72 elements that might be relevant to a given proof or theorem. This prerequisite knowledge is essential for ethical practice, as well as for deciding what to do when faced with requests to practice unethically, so ensuring learners encounter them with respect to the mathematical foundations (as well as in applications later in the learning path) is important.

2. **Identify decision-making frameworks**. The ASA GLs represent a "*virtue ethics*" decision-making framework, which can be summarized as, "what would the (ideal) ethical practitioner do in this case?" Alternatively, a *utilitarian* framework can be summarized as, "how can benefits be maximized while harms are minimized?" Either of these approaches can be useful when considering what would happen if assumptions and approximations do not hold.
3. **Identify or recognize the ethical issue**: in quantitative courses, an ethical issue can arise from failing to considering the stakeholders – and impacts on them – when assumptions and approximations do not hold.
4. **Identify and evaluate alternative actions** (on the ethical issue). There are *always* a set of decisions that can be made in any circumstance: either a) do nothing, b) consult or confer with a peer or a supervisor – using the professional guidelines or other resources, or c) report violations of policy, procedure, ethical guidelines, or law. When assumptions and approximations do not hold, the ethical practitioner is transparent, but "do nothing" would mean "not communicating that the theorem/proof is not applicable"-i.e., not being transparent. That is clearly always *an* option, but never an ethical one.
5. **Make and justify a decision**. A decision about what to do in the face of the ethical challenge that has been identified will be based on the alternatives. Justification would compare and contrast the alternatives from either the utilitarian or the virtue perspective; this is especially important if these perspectives ever differ on the course of action either supports best.
6. **Reflect on the decision**. For a course with mostly quantitative content, reflection may need to be targeted so as not to add too much narrative work (for the students to write and the instructor to grade). Examples of streamlined reflections could be: 1) ask for the most problematic assumptions/approximations of the course or section in the student's opinion, with a brief explanation of why that one is the most problematic; 2) ask students to rank order most to least problematic approximations (NB: reflection cannot be graded "right or wrong", but "do nothing" is always unethical, so questions about "how often do you think mathematics practitioners decide to "do nothing" will generate narrative that will be interesting, but might be hard to grade consistently).

Importantly, ethical reasoning can be integrated into a course, so that students can learn and practice all six aspects. However, unless the instructor also has cases or vignettes, ethical reasoning does not go beyond KSA #3, **Identify or recognize the ethical issue**. Moreover, in addition to learning the KSAs of ethical reasoning, and the ASA GLs (or similar ethical practice standard), the integration of ethical reasoning into a course requires additional material against which students can compare options in order to identify the ethical issue. For example, if the ASA Ethical Guidelines are used, then given a case or vignette, students can identify which ASA Ethical Guideline Principle(s) or elements are relevant – or have been violated – leading to the ethical issue under consideration. This can be done with a matching task, check boxes, or asking students to order a set of GL elements in terms of their relevance or importance to a violation of assumption or approximation. For eliciting narrative responses, instructors can choose a formula, proof or theorem and ask, "What decisions are made/have been made in (or in this problem's use of) this (proof/theorem)? Are assumptions, definitions, or approximations decisions? Can these affect stakeholders? If so, how; if not, why not?"

### 2.3 Stakeholder analysis: a relevant subset of Ethical Reasoning

"*Those who fund, contribute to, use, or are affected by statistical practices are considered stakeholders. The ethical statistical practitioner respects the interests of stakeholders while practicing in compliance with these Guidelines*." (ASA 2022; Principle C). As such, the impact of violations of assumptions or failures of approximations on stakeholders can be easily leveraged to integrate content about ethical statistics and data science practice into a quantitative course. Stakeholder analysis is part of Ethical Reasoning KSA #1, prerequisite knowledge (Tractenberg 2022-A, 2022-B). Importantly, stakeholder analysis can be used whether there is or is not an actual case or behavior/situation to react to, and is also independent of the use of an ethical practice standard (Tractenberg, 2022-C). Thus, a stakeholder analysis can be useful to facilitate ethical practice simply by alerting the practitioner to harms and benefits - and to which stakeholders those accrue - of any step in their workflow or analysis. Even without a specific set of ethical practice guidelines in mind, the choice that minimizes harms can be made using the stakeholder analysis. It can also be used to augment the prerequisite knowledge of a formal ethical reasoning-based case analysis (Tractenberg 2022-B).

In order to integrate a stakeholder analysis into a quantitative course, whenever an approximation or assumption is made, attention can be brought to what happens if the assumption doesn't hold, at the limits of the approximation, or the inappropriate (doesn't fit the situation, isn't right for the data, assumes more than is known/knowable) use of the definition. Table 1 (adapted from Tractenberg 2019) focuses attention on harms/benefits to create a simplified 'case analysis' that can be utilized with any formula, proof, or theorem if approximations and/or assumptions are included.

**Table 1:** Stakeholder Analysis of Harms and Benefits that accrue to seven stakeholder types

| *Stakeholder*[1]*:* | *HARM* [4,5] | *BENEFIT* [4,5] |
|---|---|---|
| YOU [2,3] | | |
| Your boss/client | | |
| Unknown individuals[2] | | |
| Employer | | |
| Colleagues | | |
| Profession | | |
| Public/public trust | | |

Notes on stakeholder analysis table:

1. Knowing to whom harms may accrue can guide learners to where the ASA GLs or other ethical practice standards can assist in decision making.
2. Articulating the harms that may accrue to YOU (the individual) is essential for following advice to "treat others' data as you would your own" (Loukides, Mason & Patil, 2018: Chapter 3). Individuals need to recognize the harms that can accrue *to them* before they can compare harms to themselves and harms to others. Moreover, "others' data" could relate to data belonging to your boss/client, your employer, unknowable others, the scientific community, or the public. Recognizing whether or not benefits or harms accrue to these different types of "others" is the only way to make a decision about how one wants other people to treat *one's* data: in someone else's table, *you* are the client, an unknowable other, or part of the public. Are some harms "worse" or some benefits "better"? Any interested instructor can augment the stakeholder analysis with additional discussion questions like these.

3. If there are no recognizable harms, and plausibly no "unknowable" harms <that would be caused by a failure of an assumption or approximation>, then there can be no conflict. It is really important to recognize whether something truly is unknowable or if it is actually something that can be known – but we/the learner just don't/doesn't know it. The key words here are "recognizable" and "plausible" – the failure to recognize something doesn't mean it does not exist. And, beware of straw man[2] or red herring[3] harms!
4. If there are plausible harms (or benefits) that cannot be identified, but are believed/suspected to exist, then there is insufficient information to make a decision and more information is needed. Recognizing this – instead of making an uninformed decision – is currently *not part of the norm.* Learning how to use this table and complete a case analysis is essential for enabling *informed decisions about ethical challenges* for *current and* future practitioners.
5. All harms are not the same; all the benefits are not the same; and harms and benefits are not exchangeable.

Whenever assumptions and/or approximations are not met, there may be impacts on stakeholders. For example, if assumptions are not met and (but) the function/algorithm works as the employer desires, a harm accrues to the employer – because they're not getting accurate information or applicable/reproducible results but a benefit may seem to accrue to the individual ("you") because the work gets completed. If the timeline is tight, the boss may also seem to benefit from an approximation being used (because it looks like the boss got the work done on time). Augmenting the discussion of assumptions and/or approximations for one construct (proof, theorem, definition, etc) can be limited to one per chapter, one per homework problem set or quiz, or one per week, depending on the ease of doing this given the content/topic. Repeating the same exercise for different proofs or assumptions/approximations will ensure that students have multiple opportunities to recognize and become competent at recognizing the impact of shortcuts and rules of thumb on stakeholders – and, offer opportunities to connect ethical practice ideas to real harms and benefits. Stakeholder analysis tables can be created with check boxes, matching, or other forms of tasks that limit the need for narrative responses from students, but still generate evaluable work products that can be graded. For eliciting narrative responses, instructors can choose one problem from each homework problem set and ask, "Who could be a stakeholder when solving a problem like (one chosen from the homework assignment)?"

### 2.4 Using the ASA Ethical Guidelines explicitly: a relevant subset of Ethical Reasoning

Combining authentic ethical content with content that is more quantitative (i.e., featuring formulae, theorems, and proofs) is difficult, but directly incorporating the ASA Ethical Guidelines (included in the Appendix) can be done without a focus on ethical reasoning or stakeholder analysis. For any given homework problem – particularly those where applications of the formula or theorem are highlighted, questions can be asked that require a matching or check-box answer (multiple choice), short answer, or narrative response. Questions can be structured in the following form:

> Choose a problem from (a homework set) and consider what would happen if, instead of following ASA Ethical Guideline Principle A, the individual didn't use appropriate methodology? Is there an application of (what you're learning) where failing to follow Principle A could negatively impact any stakeholder – including yourself?

---

[2] "Straw Man": defined as "an argument, claim or opponent that is invented in order to win or create an argument", Cambridge English Dictionary.

[3] "Red Herring": defined as "something that takes attention away from a more important subject", Cambridge English Dictionary.

This same question can be (re)formulated for all of the ASA GL Principles (A-H). Since the Appendix is focused on organizations and institutions, those items are simple to exclude from this type of question; similarly since Principle G is about leadership type roles, those might also be omitted. (Principle G and the Appendix could be leveraged to encourage students to consider attributes of the workplaces they would like to join once they graduate.) Instructors would ideally mention the GL Principles repeatedly throughout the course and other materials, and it is essential that the method by which students respond (narrative, short answer, or multiple choice) is consistent with the instructor's learning objectives for the ethical content in the course (Tractenberg 2020; Tractenberg et al. 2020).

Other GL Principle specific questions, that require the same attention to response generation and consistency with course learning objectives include:

> Choose a problem from (a homework set) and consider what would happen if, instead of following Principle D, someone didn't follow applicable rules or guidelines? Is there an application of (what you're learning) where failing to follow Principle D could negatively impact any stakeholder – including yourself? (e.g., dual use of innovation, military and <non-military>)
>
> Consider what would happen if, instead of following Principle E, someone on your working team wanted you to violate the ASA Ethical Guidelines because their profession used different Guidelines? Is there an application of (what you're learning) where failing to follow the ASA Ethical Guidelines could negatively impact any stakeholder – including yourself?
>
> Do activities that could negatively impact the profession as a stakeholder have the potential to also negatively impact yourself? Consider the alternative also: do activities that positively impact the profession as a stakeholder also positively impact yourself?

It requires effort to introduce the idea of a stakeholder into a course in an authentic way. Also there have to be actions, or reactions, that the practitioner must make, for there to be plausible stakes for people to hold (given that the "those who fund, contribute to, use, *or are affected by statistical practices* are considered stakeholders"). This is why we focus on the effects of failures of assumptions or approximations to hold as the point at which purely quantitative content (formula, proof, theorem) can begin to bear the integration of ethical content. These failures or violations require attention to both the formula, proof, or theorem (i.e., the actual target content of the course or lesson) and also to the possible harms or benefits that the otherwise purely quantitative content can have on stakeholders.

## 3. Integrating ethical content in a Mathematical Statistics course

An undergraduate course in Mathematical Statistics - typically the first exposure for students in the major to the "underpinnings" of statistical inference - has been revised for the 2022-23 Fall term to integrate ethical practice content. This was undertaken by one of us (ST) with input from the other (RET). This section presents the outline of the syllabus (without timings so the reader can map to their own term (summer session or quarter/semester). Subsequent sections discuss the features and decisions represented in the syllabus, a fuller version of which is incliuded in the Appendix.

### 3.1 Syllabus outline

Learning objectives (referred to throughout the syllabus)

1. Understand frequentist and Bayesian methods for parameter estimation and common approaches to evaluate and compare estimators.
2. Work with asymptotic theorems to characterize the behavior of common types of estimators.
3. Understand how to analytically derive a Bayesian posterior distribution and how to interpret a Bayesian credible interval and demonstrate familiarity with different types of priors.
4. Understand the important role of likelihood functions for hypothesis testing in both Bayesian and frequentist frameworks
5. Understand the relationship between frequentist confidence interval estimation and hypothesis testing.
6. Familiarity with common types of optimal testing strategies.
7. Construct and interpret frequentist p-values for hypothesis tests and interpret the error rates.
8. Identify/define model parameters and state statistical inferential questions in terms of these parameters for various common, realistic study settings.
9. **Contextualize statistical methods and theory in science and policy at large and develop a habit of mind informed by how the application of such methods can affect stakeholders.**
10. **Understand what it means to be a steward of statistics, how to be stewardly, and why that is important.**[4]

Course Content and Order of Topics Introduced:

Unit One - **Thoughtful use of**[5] estimation techniques

Relevant Learning Objectives: 1, 2, 3, 9

1.1 Review of probability
1.2 Methods to derive estimators
1.3 Intro to large sample theory of estimation
1.4 Interval estimation

---

[4] Two learning outcomes reflect the instructor's intention to prioritize not only the contextualization of statistical methods and theory in real world applications, but also the facts that a) statistical methods and theories have potential impacts on stakeholders; b) these impacts and stakeholders are important enough to feature at the learning outcome level (not as an aside); and c) there is more to learning the mathematical underpinnings of statistics than "just math" - stewardship, and responsibilities to the discipline, are also learnable and important to internalize.

[5] The addition of "**Thoughtful use of**" to the Unit title signals to students that the focus moves beyond "use" and learners will be expected to be thoughtful (not simply memorizing) in their selection and application of estimation techniques.

Example activities for Unit One:

- Read *Cargo cult statistics and the scientific crisis* (Stark and Saltelli, 2016) before class and in-class discussion on what students found surprising and why.
- Groups instructed to list as many stakeholders as students can identify in the classroom and, hypothetically, in the workplace where they may use statistics.
- Instructor led stakeholder analysis regarding the application of an estimator with desirable large sample properties in the development of a drug.

Unit Two - Statistical inference **and stewardship**[6]
Relevant Learning Objectives: 4, 5, 6, 7, 10

2.1 What's in a likelihood
2.2 Neyman-Pearson paradigm for optimal testing
2.3 Trustworthy testing

Example activities for Unit Two:

- Read blurb from *The preparation of stewards with the Mastery Rubric for Stewardship: Re-envisioning the formation of scholars and practitioners* (Rios, Golde & Tractenberg; 2019) before class followed by in-class, small group discussion about the meaning of the phrase "steward of statistics".
- Individual reflection on what it means to be a steward of statistical practices handed in as an ungraded assignment that will be revisited later in the semester.
- Student groups complete an in-class worksheet identifying potential stakeholders in the application of hypothesis testing.
- Individual assessment item that tasks students with making a decision from a hypothesis test that is justified with a stakeholder analysis.

Unit Three - **Disciplinary best practices**[7] for common study designs and analyses
Relevant Learning Objectives: 8, 9, 10

3.1 One and two sample means
3.2 ANOVA
3.3 Categorical data
3.4 Linear regression

Example activities for Unit 3:

- Student groups assigned various principles from the American Statistical Association's Guidelines for Ethical Statistical Practice (Ie. Principle A through H) complete a worksheet where they identify which principles specifically they think are most relevant for statistical estimation and which are most relevant for inference. Followed by an instructor-led discussion where the considerations for estimation and inference are compared and contrasted.

---

[6] The addition of "**and stewardship**" to the Unit title signals to students that the focus on inference includes stewardly consideration.

[7] The addition of "**disciplinary best practices**" for common designs and analyses informs students that respect for the discipline (i.e., stewardship) is equally important for them to learn as the designs and analytic methods. The Unit title transmits the instructor's intention that students develop a stewardly, discipline- rather than results-centered, mindset.

- In-class group activity to identify which Ethical Principles are most relevant for each statistical method (i.e.two sample inference, ANOVA models, inference about categorical data, and inference with linear regression models).
- Individual reflection on what the phrase "steward of statistics" means to each student.
- Individual reflection on students' projected growth as a steward of statistical practice incorporated into end of course evaluation. (Students' are sent a copy of their anonymous responses.)

**3.2. Notes on syllabus elements: Unit One. Motivate and contextualize**

The first unit of the semester serves to both motivate and provide language for ethical reasoning in statistics. The content is focused on assuring student facilities with the prerequisite knowledge from their prior coursework in mathematics and statistics. In Week 1, students are asked to read *Cargo cult statistics and the scientific crisis* (Stark and Saltelli, 2016). This article presents an approachable argument for statistical practitioners to carefully consider potential consequences of their statistical choices, particularly on the scientific crisis and subsequent denigrations to the credibility of science generally. Students become aware of potential stakeholders though they may not yet have the language to identify them as such.

In previous academic years the syllabus has not featured mention of ethical content (such as "stewardship" and "best practice"), but now it not only is featured in the syllabus but ethics-relevant discussions also begin in the earliest lectures. The syllabus above communicates an intention to inculcate a sense of professional identity, ethical obligations to respect stakeholders, and the idea that "even the mathematical underpinnings" of statistics accrue responsibilities (e.g. assuring that approximations and assumptions are plausible). To optimize the chances of meeting learning objectives 9 and 10, this integration needs to begin at the earliest point in the term and carry through.

A group activity over subsequent weeks could involve asking students to identify stakeholders in different vignettes. In addition to recognizing stakeholders in diverse contexts, students are also encouraged to identify diverse stakeholders, for example, within the classroom itself. This prompts learners to expand their perspective beyond a generic teacher-student relationship to promote the realization that their peers and maybe even the entire department are stakeholders to consider. Students can then begin to practice shifting their perspectives while they are in class. This permits exploration along many, possibly under-appreciated, lines of questioning about the statistical material which, in this unit, is primarily a mathematical introduction to estimation. In-class and homework activities are intended to familiarize students with tools and their respective properties; introduce considerations of, and beyond, mathematical characteristics in estimation; and practice shifting perspectives and developing some fluency about these quantitative ideas through discussion/communication with peers.

Near the end of the unit, once the statistical reasoning and mathematical mechanics behind selecting estimators has been practiced (in homeworks) and assessed (in quizzes), the instructor presents a specific vignette in which to consider an estimation decision. For example, having learned about the general conditions under which the distribution of a maximum likelihood estimator converges to a Gaussian distribution, consider the use of an asymptotic confidence interval to estimate any moment or parameter. Any vignette can be created, with the overall intention of encouraging the realization that statistical and mathematical properties of estimators are theoretically helpful, but cannot mitigate poor design or weak/incorrect understanding of population/sample.

**3.3 Notes on syllabus element**s: **Unit Two. From examples to practice**

Moving into the second unit, students have been introduced to the concept of stakeholders and have some familiarity with identifying, and strategically assessing, the pros and cons of statistical decisions through a stakeholder analysis. Now as they are introduced to the concept of statistical hypothesis testing, they will also spend some time continuing to build their fluency while considering specific ethical guidance, stewardly responsibilities, and stakeholder impacts that are related to inference. The steward is the individual to whom "we can entrust the vigor, quality, and integrity of the field" (Golde & Walker, 2006: p. 5), and is someone who "will creatively generate new knowledge, critically conserve valuable and useful ideas, and responsibly transform those understandings through writing, teaching, and application" (Golde & Walker, 2006: p. 5). Rios, Golde & Tractenberg (2019) articulated the knowledge, skills, and abilities of the disciplinary and professional steward, extending the original definition (from describing the doctorate holder in 2006) to describing any practitioner in a field or discipline. In order to make stewardship into something that was learnable and improvable, like ethical reasoning, a unique stewardship KSA was developed "…to describe the responsibility to recognize when these behaviors need to be applied or modeled... We have called this KSA "requisite knowledge/situational awareness" to capture the attention that would be given to standards of professional practice (if they exist) during education or training, or when orienting new employees in the workplace." (Rios et al. p. 11 of 27). Understanding the construct of stewardship provides a frame of reference for the student, in terms of understanding that statistical practitioners incur obligations to stakeholders because the stewardly practitioner is entrusted by these stakeholders to practice ethically. Outside of class, students can be asked to read sections of Rios, Golde & Tractenberg (2019), for example, followed by in-class small group discussions about what the phrase "steward of statistics" means to them upon their introduction to the concept of stewardship.

Having the ASA Ethical Guidelines (2022) as a reference reinforces to students that "statistical practice" is much more than simply choosing or obtaining a data set and a software program, and combining those together. There are 8 Principles and an Appendix, with a total of 72 elements. These are not valuable if they are simply memorized, so the activities of reading through the Guidelines to identify all the elements that might pertain, or that would be followed by a stewardly practitioner, can help to initiate the development of a sense of professional identity. Within the context of hypothesis testing, students learn the extent to which automated software can/can not assist in making decisions. In class, students will practice using a stakeholder analysis to explore the intricacies in formulating a hypothesis to test, in addition to assessing the fallout of various errors. Considering a simple analysis with two stakeholders, the statistical practitioner and their employer, groups can be assigned to assess potential harms and benefits for, say, the choice of null and alternative, the decision to focus on power or error rates, using a p-value to make a decision, or using a Bayesian posterior to make a decision.

A group activity over subsequent weeks is to engage in discussions, or include the identification, of ASA principles that are relevant for hypothesis testing (in general), and for the selection, use, and reporting of results of hypothesis testing methods. These can be in multiple choice and short answer formats (e.g., list the Principle and specific elements that are relevant; and/or give a brief explanation about choosing one or another Principle or element). Additionally, student engagement with the stakeholder analysis can continue and become more advanced. For example, homework or in-class problems can request stakeholder analysis for the choice of a hypothesis testing method, or its use to make decisions or draw conclusions.

In previous academic years, the statistics course syllabi had not featured the ASA Ethical Guidelines. Instead, roughly 30 minutes of class time was dedicated to an activity that introduced students to the Ethical Guidelines for Principle B: Integrity of Data and Methods. First, the instructor presented these guidelines to the students, having a different person read each principle aloud. Then, students were given time to individually reflect on a real application of statistics which, unbeknownst to them, was an application with known ethical violations. During the reflection, students were prompted to identify which guidelines may have been violated and to consider what additional information they may need to determine if any ethical violations had occurred. Finally, the exercise concluded with the instructor revealing the known issues in the statistical analysis and provided some historical context as to how these issues were discovered. This activity occurred later in the semester, once students had been introduced to specific statistical methods for testing and estimation. By contrast, in the syllabus presented here, there are repeated opportunities for students to revisit the Ethical Guidelines and these are couched within a larger framework of stewardship. The intentions are: a) to focus student attention on these features of ethical practice; b) to get students to engage with various aspects of "statistical practice" that should be done ethically; and c) to allow students sufficient time to consider and work with the Ethical Guidelines, that can support development of the desired habits of mind that have the strongest likelihood of enduring beyond the end of the course. Practically speaking, at the end of the second unit, students will have begun to develop an ethically informed schema for answering questions about data with statistics. At this stage in the term, students are presented with a framework through the concept of stewardship though they may not yet understand how this concept could be relevant for them as potential statistical practitioners in the future.

### 3.4 Notes on syllabus elements: Unit Three. Independent and collaborative critical thinking

This semester-long course focuses on the mathematical and contextual considerations that are critical for effective statistical practice. The intention throughout the syllabus and course is to emphasize that effective statistical practice is more than just the application of mathematical principles, or the use of software or execution of analysis. Instead, critical thinking is required throughout statistical practice as reflected in the learning objectives of the course. During the final third period of the term, students are regularly tasked with synthesizing what they have learned. They will apply the knowledge they have accumulated regarding estimation, parameter and method selection, and the uses of this knowledge in testing hypotheses and drawing inference while also keeping in mind stakeholders, the ethical practice standards of the ASA, and how these should be considered by the stewardly practitioner. The syllabus is designed so that the first two thirds of the term focuses on training students for the practice of regular, critical mathematical reasoning both independently and in collaboration with their peers. The new techniques and methods students are learning in the last third period of the term are supported by specific learning objectives which are consistently revisited during regular assessments and group activities. Moreover, homework and in-class problems involving the application of these methods can now require justification for decision making in the form of a stakeholder analysis for (any) decisions made along the way from study design, to data collection, analysis, and interpretation. Assessments can also prompt students to identify any ASA Ethical Guideline elements that are relevant along this same path (from design to interpretation).

As the term concludes, students are asked again to consider the importance of the stewardly practice of statistics and given the opportunity to compare their answers to their responses from earlier in the term when they were first introduced to the concept of stewardship. Finally, students can be asked to describe how they anticipate stakeholder analysis, the ASA Ethical Guidelines, and their sense of themselves as stewards to continue to grow in future courses. This simple activity can promote a commitment to ongoing engagement with ethical practice throughout their educational training and can be readily incorporated into end of semester course evaluations, for example.

In previous academic years the introduction of ethical content was limited to simply familiarizing the students with the existence of the ASA Ethical Guidelines and a single case study of an inappropriate application of statistical analysis. In the current syllabus however, the entire third unit features active integration of thinking about stakeholders, stewardship, and how the ASA Ethical Guidelines can facilitate both. More than familiarization with disciplinary best practices is required in this version of a statistics course syllabus. From the first day of class when the syllabus is introduced, students are made aware that an explicit objective for their training in the course is to develop new habits of mind that are formed by consideration of stakeholders in statistical practice and a sense of professional identity based on the notion of stewardship. They are introduced to new concepts in stages that align with the presentation of the typical statistical course material. Ethical reasoning for statistical practice is scaffolded into the course material so that it is not until after students have observed and practiced the stakeholder analysis and professional guidelines for stewardship of the discipline that they are tasked with reasoning through any particular example of statistical analysis in a real world problem. This structured approach, rather than a one-off detour through a single case analysis for example, improves the likelihood that the last two learning objectives are actually met.

## 4. Discussion and Conclusions

### 4.1 Discussion

We have outlined the rationale behind three different approaches to integrating ethical content into a primarily-quantitative course. The example course, Mathematical Statistics, is described in terms of an adaptable syllabus. The syllabus includes explicit signaling from the instructor as to the importance and role of stewardship, stakeholders, and the ASA Ethical Guidelines for Statistical Practice. These three dimensions, particularly as they are integrated into the course as discussed, represent the first element of ethical reasoning: prerequisite knowledge. Just as with the mathematics and statistics content, the prerequisite knowledge required for ethical reasoning and ethical statistical practice goes beyond memorization. Decisions that the syllabus implicitly reflects include a focus on solely the prerequisite knowledge dimension of the ethical reasoning paradigm (Tractenberg & FitzGerald, 2012; Tractenberg, 2022-A; 2022-B); and a focus on the construct of stewardship, without moving through its set of knowledge, skills, and abilities (Rios et al. 2019). The ethical content included in this course, as described with the syllabus we have included, could be built upon in later courses with the same types of activities, but with more advanced problems or responses by students. Or, programs that use the integration described in this syllabus can choose ethical reasoning (Tractenberg & FitzGerald, 2012; Tractenberg et al. 2017) or stewardship (Rios et al. 2019) as a paradigm to graft into the full curriculum.

### 4.2 Conclusions

Recent encouragement to incorporate "ethics" into the curricula for statistics (ASA, 2014) and for data science (National Academies, 2018) has been published for undergraduate degree programs. However, resources that exist for "teaching ethics" are not accessible for many instructors of courses that feature primarily quantitative materials and content. When discussing the best practices for teaching mathematics, Kelton (2010) notes,

> "It is easy for students to have the misconception that the problems are easy when they witness you solving them with no difficulty. Giving the students a chance to see where they might stumble will help them formulate questions and motivate them to give the homework proper attention. It may be best to only give the students one or two problems at a time to solve on their own. That is generally sufficient for the students to ascertain whether they are having significant difficulty. After you have demonstrated the skill, give the class a basic problem or two to try on their own."

Not only is this sentiment as true for reasoning and understanding ethical obligations as it is for teaching mathematics, it is also appropriate for instructors who want to integrate ethics content into more mathematically-oriented courses. In their Recommendation 2.5, The Committee on Envisioning the Data Science Discipline pointed out that "Ethics is a topic that, given the nature of data science, students should learn and practice throughout their education. Academic institutions should ensure that ethics is woven into the data science curriculum from the beginning and throughout." (National Academies, 2018). This paper has outlined three different ways to overcome the difficulties that might prevent instructors in primarily quantitative courses in these curricula from integrating ethical content, ethical reasoning, or the consideration of how fundamental attributes of formulae, theorems, and proofs can support this integration "from the beginning" of quantitative programs or courses. A syllabus and organizational structure is presented that can be used, adapted, or utilized in - or to seed - a sequence of courses where students are asked to generate increasingly sophisticated responses to similar questions.

## Acknowledgements

Suzanne Thornton acknowledges support from Swarthmore College Faculty Research Support Grants.

# APPENDIX: Full Syllabus and ASA Ethical Guidelines for Statistical Practice

**Syllabus for a First Course in Mathematical Statistics**

**Learning objectives**

1. Understand frequentist and Bayesian methods for parameter estimation and common approaches to evaluate and compare estimators.
2. Work with asymptotic theorems to characterize the behavior of common types of estimators.
3. Understand how to analytically derive a Bayesian posterior distribution and how to interpret a Bayesian credible interval and demonstrate familiarity with different types of priors.
4. Understand the important role of likelihood functions for hypothesis testing in both Bayesian and frequentist frameworks
5. Understand the relationship between frequentist confidence interval estimation and hypothesis testing.
6. Familiarity with common types of optimal testing strategies.
7. Construct and interpret frequentist p-values for hypothesis tests and interpret the error rates.
8. Identify/define model parameters and state statistical inferential questions in terms of these parameters for various common, realistic study settings.
9. **Contextualize statistical methods and theory in science and policy at large and develop a habit of mind informed by how the application of such methods can affect stakeholders.**
10. **Understand what it means to be a steward of statistics, how to be stewardly, and why that is important[8].**

**Topics outline for the semester**

**Unit One - Thoughtful use of[9] estimation techniques**

Relevant Learning Objectives: 1, 2, 3, 9

---

[8] Two learning outcomes reflect the instructor's intention to prioritize not only the contextualization of statistical methods and theory in real world applications, but also the facts that a) statistical methods and theories have potential impacts on stakeholders; b) these impacts and stakeholders are important enough to feature at the learning outcome level (not as an aside); and c) there is more to learning the mathematical underpinnings of statistics than "just math" - stewardship, and responsibilities to the discipline, are also learnable and important to internalize.

[9] The addition of "**Thoughtful use of**" to the Unit title signals to students that the focus moves beyond "use" and learners will be expected to be thoughtful (not simply memorizing) in their selection and application of estimation techniques.

*1.1 Review of probability*
Throughout the review of probability, the instructor will introduce a reasoning tool called a "stakeholder analysis". Students will practice naming stakeholders in various contexts and begin to understand that although assumptions and approximations always work in the theoretical realm, they have real implications for practice and may not work for every stakeholder.

*1.2 Methods to derive estimators*
Throughout the introduction of estimation methods, the instructor will continue to feature the stakeholder analysis to explore how selection and application of estimation techniques can potentially have effects beyond simple numeric solutions.

*1.3 Intro to large sample theory of estimation*
After introducing the students to fundamental large sample results concerning maximum likelihood estimation, the instructor will lead an in-class discussion of a stakeholder analysis regarding the application of an estimator with desirable large sample properties.

*1.4 Interval estimation*
Students are introduced to Bayesian and frequentist interval estimation techniques. The instructor will lead a class discussion of a stakeholder analysis focused on the implications of various choices that are made in interval estimation.

**Unit Two - Statistical inference and stewardship**[10]
Relevant Learning Objectives: 4, 5, 6, 7, 10

*2.1 What's in a likelihood*
As students' work through understanding the concept of a likelihood function and dimension reduction through sufficient statistics, the instructor will present the notion of a steward of statistics.

*2.2 Neyman-Pearson paradigm for optimal testing*
As students' wrestle with the logic behind optimal testing, they are also becoming aware of the restrictiveness of ideal mathematical reasoning in the applied world. They will be encouraged to reference an explicit definition of stewardship as they consider how to use statistical tests in less-than-ideal conditions.

*2.3 Trustworthy testing*
As students start to understand what it means to be a steward of statistical practice, the instructor will reference hypothesis tests as a concrete example of why stewardship is important. The instructor will lead students to examine the role of various potential stakeholders in stewardly statistical inference.

**Unit Three - Disciplinary best practices for**[11] **common study designs and analyses**
Relevant Learning Objectives: 8, 9, 10

---

[10] The addition of "**and stewardship**" to the Unit title signals to students that the focus on inference includes stewardly consideration.
[11] The addition of "**Disciplinary best practices for**" common designs and analyses informs students that respect for the discipline (i.e., stewardship) is equally important for them to learn as the designs and analytic methods. The Unit title transmits the instructor's intention that students develop a stewardly, discipline- rather than results-centered, mindset.

*3.1 One and two sample means*
The instructor will introduce the American Statistical Association's Guidelines for Ethical Statistical Practice. Students will work in small discussion groups to consider why particular statistical methods are “common” from a perspective that names and considers potential stakeholders.

*3.2 ANOVA*
The instructor will continue to reference the ASA's Ethical Guidelines with respect to ANOVA
methods. Groups will continue to attempt to characterize "common" statistical methods and will be encouraged to also consider the role of software and the level of engagement with ASA Ethical Guidelines that software encourages.

*3.3 Categorical data*
Students will be prompted to discuss the differences for stakeholders between methods that use continuous or categorical data. The instructor will emphasize what the ASA's Ethical Guidelines suggest a stewardly practitioner should prioritize when given a choice between categorical or continuous variable-based analysis.

*3.4 Linear regression*
As the semester concludes, students revisit the concept of linear regression from a more mathematically and stewardly informed perspective. In groups and as a class, students and the instructor will discuss variable choice and model selection referencing the ASA's Ethical Guidelines and centering the notion of statistical stewardship.

## Ethical Guidelines for Statistical Practice

***Prepared by the Committee on Professional Ethics***
***of the American Statistical Association***
**31 January 2022**

### PURPOSE OF THE GUIDELINES:

The American Statistical Association's Ethical Guidelines for Statistical Practice are intended to help statistical practitioners make decisions ethically. In these Guidelines, “statistical practice” includes activities such as: designing the collection of, summarizing, processing, analyzing, interpreting, or presenting, data; as well as model or algorithm development and deployment. Throughout these Guidelines, the term "statistical practitioner" includes all those who engage in statistical practice, regardless of job title, profession, level, or field of degree. The Guidelines are intended for individuals, but these principles are also relevant to organizations that engage in statistical practice.

The Ethical Guidelines aim to promote accountability by informing those who rely on any aspects of statistical practice of the standards that they should expect. Society benefits from informed judgments supported by ethical statistical practice. All statistical practitioners are expected to follow these Guidelines and to encourage others to do the same.

In some situations, Guideline principles may require balancing of competing interests. If an unexpected ethical challenge arises, the ethical practitioner seeks guidance, not exceptions, in the Guidelines. To justify unethical behaviors, or to exploit gaps in the Guidelines, is unprofessional, and inconsistent with these Guidelines.

## PRINCIPLE A: Professional Integrity and Accountability

Professional integrity and accountability require taking responsibility for one's work. Ethical statistical practice supports valid and prudent decision making with appropriate methodology. The ethical statistical practitioner represents their capabilities and activities honestly, and treats others with respect.

**The ethical statistical practitioner:**

1. Takes responsibility for evaluating potential tasks, assessing whether they have (or can attain) sufficient competence to execute each task, and that the work and timeline are feasible. Does not solicit or deliver work for which they are not qualified, or that they would not be willing to have peer reviewed.
2. Uses methodology and data that are valid, relevant, and appropriate, without favoritism or prejudice, and in a manner intended to produce valid, interpretable, and reproducible results.
3. Does not knowingly conduct statistical practices that exploit vulnerable populations or create or perpetuate unfair outcomes.
4. Opposes efforts to predetermine or influence the results of statistical practices, and resists pressure to selectively interpret data.
5. Accepts full responsibility for their own work; does not take credit for the work of others; and gives credit to those who contribute. Respects and acknowledges the intellectual property of others.
6. Strives to follow, and encourages all collaborators to follow, an established protocol for authorship. Advocates for recognition commensurate with each person's contribution to the work. Recognizes that inclusion as an author does imply, while acknowledgement may imply, endorsement of the work.
7. Discloses conflicts of interest, financial and otherwise, and manages or resolves them according to established policies, regulations, and laws.
8. Promotes the dignity and fair treatment of all people. Neither engages in nor condones discrimination based on personal characteristics. Respects personal boundaries in interactions and avoids harassment including sexual harassment, bullying, and other abuses of power or authority.
9. Takes appropriate action when aware of deviations from these Guidelines by others.
10. Acquires and maintains competence through upgrading of skills as needed to maintain a high standard of practice.
11. Follows applicable policies, regulations, and laws relating to their professional work, unless there is a compelling ethical justification to do otherwise.
12. Upholds, respects, and promotes these Guidelines. Those who teach, train, or mentor in statistical practice have a special obligation to promote behavior that is consistent with these Guidelines.

## PRINCIPLE B: Integrity of Data and Methods

The ethical statistical practitioner seeks to understand and mitigate known or suspected limitations, defects, or biases in the data or methods and communicates potential impacts on the interpretation, conclusions, recommendations, decisions, or other results of statistical practices.

**The ethical statistical practitioner:**

1. Communicates data sources and fitness for use, including data generation and collection processes and known biases. Discloses and manages any conflicts of interest relating to the data sources. Communicates data processing and transformation procedures, including missing data handling.
2. Is transparent about assumptions made in the execution and interpretation of statistical practices including methods used, limitations, possible sources of error, and algorithmic biases. Conveys results or applications of statistical practices in ways that are honest and meaningful.
3. Communicates the stated purpose and the intended use of statistical practices. Is transparent regarding a priori versus post hoc objectives and planned versus unplanned statistical practices. Discloses when multiple comparisons are conducted, and any relevant adjustments.
4. Meets obligations to share the data used in the statistical practices, for example, for peer review and replication, as allowable. Respects expectations of data contributors when using or sharing data. Exercises due caution to protect proprietary and confidential data, including all data that might inappropriately harm data subjects.
5. Strives to promptly correct substantive errors discovered after publication or implementation. As appropriate, disseminates the correction publicly and/or to others relying on the results.
6. For models and algorithms designed to inform or implement decisions repeatedly, develops and/or implements plans to validate assumptions and assess performance over time, as needed. Considers criteria and mitigation plans for model or algorithm failure and retirement.
7. Explores and describes the effect of variation in human characteristics and groups on statistical practice when feasible and relevant.

## PRINCIPLE C: Responsibilities to Stakeholders

Those who fund, contribute to, use, or are affected by statistical practices are considered stakeholders. The ethical statistical practitioner respects the interests of stakeholders while practicing in compliance with these Guidelines.

**The ethical statistical practitioner:**

1. Seeks to establish what stakeholders hope to obtain from any specific project. Strives to obtain sufficient subject-matter knowledge to conduct meaningful and relevant statistical practice.
2. Regardless of personal or institutional interests or external pressures, does not use statistical practices to mislead any stakeholder.
3. Uses practices appropriate to exploratory and confirmatory phases of a project, differentiating findings from each so the stakeholders can understand and apply the results.

4. Informs stakeholders of the potential limitations on use and re-use of statistical practices in different contexts and offers guidance and alternatives, where appropriate, about scope, cost, and precision considerations that affect the utility of the statistical practice.
5. Explains any expected adverse consequences from failing to follow through on an agreed-upon sampling or analytic plan.
6. Strives to make new methodological knowledge widely available to provide benefits to society at large. Presents relevant findings, when possible, to advance public knowledge.
7. Understands and conforms to confidentiality requirements for data collection, release, and dissemination and any restrictions on its use established by the data provider (to the extent legally required). Protects the use and disclosure of data accordingly. Safeguards privileged information of the employer, client, or funder.
8. Prioritizes both scientific integrity and the principles outlined in these Guidelines when interests are in conflict.

## PRINCIPLE D: Responsibilities to Research Subjects, Data Subjects, or those directly affected by statistical practices

The ethical statistical practitioner does not misuse or condone the misuse of data. They protect and respect the rights and interests of human and animal subjects. These responsibilities extend to those who will be directly affected by statistical practices.

**The ethical statistical practitioner:**

1. Keeps informed about and adheres to applicable rules, approvals, and guidelines for the protection and welfare of human and animal subjects. Knows when work requires ethical review and oversight.[12]
2. Makes informed recommendations for sample size and statistical practice methodology in order to avoid the use of excessive or inadequate numbers of subjects and excessive risk to subjects
3. For animal studies, seeks to leverage statistical practice to reduce the number of animals used, refine experiments to increase the humane treatment of animals, and replace animal use where possible.
4. Protects people's privacy and the confidentiality of data concerning them, whether obtained from the individuals directly, other persons, or existing records. Knows and adheres to applicable rules, consents, and guidelines to protect private information.
5. Uses data only as permitted by data subjects' consent when applicable or considering their interests and welfare when consent is not required. This includes primary and secondary uses, use of repurposed data, sharing data, and linking data with additional data sets.
6. Considers the impact of statistical practice on society, groups, and individuals. Recognizes that statistical practice could adversely affect groups or the public perception of groups, including marginalized groups. Considers approaches to minimize negative impacts in applications or in framing results in reporting.
7. Refrains from collecting or using more data than is necessary. Uses confidential information only when permitted and only to the extent necessary. Seeks to minimize the risk of re-identification when sharing de-identified data or results where there is an expectation of confidentiality. Explains any impact of de-identification on accuracy of results.

[12] Examples of ethical review and oversight include an Institutional Review Board (IRB), an Institutional Animal Care and Use Committee (IACUC), or a compliance assessment.

8. To maximize contributions of data subjects, considers how best to use available data sources for exploration, training, testing, validation, or replication as needed for the application. The ethical statistical practitioner appropriately discloses how the data are used for these purposes and any limitations.
9. Knows the legal limitations on privacy and confidentiality assurances and does not over-promise or assume legal privacy and confidentiality protections where they may not apply.
10. Understands the provenance of the data, including origins, revisions, and any restrictions on usage, and fitness for use prior to conducting statistical practices.
11. Does not conduct statistical practice that could reasonably be interpreted by subjects as sanctioning a violation of their rights. Seeks to use statistical practices to promote the just and impartial treatment of all individuals.

## PRINCIPLE E: Responsibilities to members of multidisciplinary teams

Statistical practice is often conducted in teams made up of professionals with different professional standards. The statistical practitioner must know how to work ethically in this environment.

**The ethical statistical practitioner:**

1. Recognizes and respects that other professions may have different ethical standards and obligations. Dissonance in ethics may still arise even if all members feel that they are working towards the same goal. It is essential to have a respectful exchange of views.
2. Prioritizes these Guidelines for the conduct of statistical practice in cases where ethical guidelines conflict.
3. Ensures that all communications regarding statistical practices are consistent with these Guidelines. Promotes transparency in all statistical practices.
4. Avoids compromising validity for expediency. Regardless of pressure on or within the team, does not use inappropriate statistical practices.

## PRINCIPLE F: Responsibilities to Fellow Statistical Practitioners and the Profession

Statistical practices occur in a wide range of contexts. Irrespective of job title and training, those who practice statistics have a responsibility to treat statistical practitioners, and the profession, with respect. Responsibilities to other practitioners and the profession include honest communication and engagement that can strengthen the work of others and the profession.

**The ethical statistical practitioner:**

1. Recognizes that statistical practitioners may have different expertise and experiences, which may lead to divergent judgments about statistical practices and results.

Constructive discourse with mutual respect focuses on scientific principles and methodology and not personal attributes.

2. Helps strengthen, and does not undermine, the work of others through appropriate peer review or consultation. Provides feedback or advice that is impartial, constructive, and objective.
3. Takes full responsibility for their contributions as instructors, mentors, and supervisors of statistical practice by ensuring their best teaching and advising -- regardless of an academic or non-academic setting -- to ensure that developing practitioners are guided effectively as they learn and grow in their careers.
4. Promotes reproducibility and replication, whether results are "significant" or not, by sharing data, methods, and documentation to the extent possible.
5. Serves as an ambassador for statistical practice by promoting thoughtful choices about data acquisition, analytic procedures, and data structures among non-practitioners and students. Instills appreciation for the concepts and methods of statistical practice.

## PRINCIPLE G: Responsibilities of Leaders, Supervisors, and Mentors in Statistical Practice

Statistical practitioners leading, supervising, and/or mentoring people in statistical practice have specific obligations to follow and promote these Ethical Guidelines. Their support for – and insistence on – ethical statistical practice are essential for the integrity of the practice and profession of statistics as well as the practitioners themselves.

**Those leading, supervising, or mentoring statistical practitioners are expected to**:

1. Ensure appropriate statistical practice that is consistent with these Guidelines. Protect the statistical practitioners who comply with these Guidelines, and advocate for a culture that supports ethical statistical practice.
2. Promote a respectful, safe, and productive work environment. Encourage constructive engagement to improve statistical practice.
3. Identify and/or create opportunities for team members/mentees to develop professionally and maintain their proficiency.
4. Advocate for appropriate, timely, inclusion and participation of statistical practitioners as contributors/collaborators. Promote appropriate recognition of the contributions of statistical practitioners, including authorship if applicable.
5. Establish a culture that values validation of assumptions, and assessment of model/algorithm performance over time and across relevant subgroups, as needed. Communicate with relevant stakeholders regarding model or algorithm maintenance, failure, or actual or proposed modifications.

## PRINCIPLE H: Responsibilities Regarding Potential Misconduct

The ethical statistical practitioner understands that questions may arise concerning potential misconduct related to statistical, scientific, or professional practice. At times, a practitioner may accuse someone of misconduct, or be accused by others. At other times, a practitioner may be involved in the investigation of others' behavior. Allegations of misconduct may arise within different institutions with different standards and potentially different outcomes. The elements that follow relate specifically to allegations of statistical, scientific, and professional misconduct.

**The ethical statistical practitioner:**

1. Knows the definitions of, and procedures relating to, misconduct in their institutional setting. Seeks to clarify facts and intent before alleging misconduct by others. Recognizes that differences of opinion and honest error do not constitute unethical behavior.
2. Avoids condoning or appearing to condone statistical, scientific, or professional misconduct. Encourages other practitioners to avoid misconduct or the appearance of misconduct.
3. Does not make allegations that are poorly founded, or intended to intimidate. Recognizes such allegations as potential ethics violations.
4. Lodges complaints of misconduct discreetly and to the relevant institutional body. Does not act on allegations of misconduct without appropriate institutional referral, including those allegations originating from social media accounts or email listservs.
5. Insists upon a transparent and fair process to adjudicate claims of misconduct. Maintains confidentiality when participating in an investigation. Discloses the investigation results honestly to appropriate parties and stakeholders once they are available.
6. Refuses to publicly question or discredit the reputation of a person based on a specific accusation of misconduct while due process continues to unfold.
7. Following an investigation of misconduct, supports the efforts of all parties involved to resume their careers in as normal a manner as possible, consistent with the outcome of the investigation.
8. Avoids, and acts to discourage, retaliation against or damage to the employability of those who responsibly call attention to possible misconduct.

## GUIDELINES APPENDIX
### Responsibilities of organizations/institutions

Whenever organizations and institutions design the collection of, summarize, process, analyze, interpret, or present, data; or develop and/or deploy models or algorithms, they have responsibilities to use statistical practice in ways that are consistent with these Guidelines, as well as promote ethical statistical practice.

**Organizations and institutions engage in, and promote, ethical statistical practice by**:

1. Expecting and encouraging all employees and vendors who conduct statistical practice to adhere to these Guidelines. Promoting a workplace where the ethical practitioner may apply the Guidelines without being intimidated or coerced. Protecting statistical practitioners who comply with these Guidelines.
2. Engaging competent personnel to conduct statistical practice, and promote a productive work environment.
3. Promoting the professional development and maintenance of proficiency for employed statistical practitioners.
4. Supporting statistical practice that is objective and transparent. Not allowing organizational objectives or expectations to encourage unethical statistical practice by its employees.

5. Recognizing that the inclusion of statistical practitioners as authors, or acknowledgement of their contributions to projects or publications, requires their explicit permission because it may imply endorsement of the work.
6. Avoiding statistical practices that exploit vulnerable populations or create or perpetuate discrimination or unjust outcomes. Considering both scientific validity and impact on societal and human well-being that results from the organization's statistical practice.
7. Using professional qualifications and contributions as the basis for decisions regarding statistical practitioners' hiring, firing, promotion, work assignments, publications and presentations, candidacy for offices and awards, funding or approval of research, and other professional matters.

**Those in leadership, supervisory, or managerial positions who oversee statistical practitioners promote ethical statistical practice by following Principle G and:**

8. Recognizing that it is contrary to these Guidelines to report or follow only those results that conform to expectations without explicitly acknowledging competing findings and the basis for choices regarding which results to report, use, and/or cite.
9. Recognizing that the results of valid statistical studies cannot be guaranteed to conform to the expectations or desires of those commissioning the study or employing/supervising the statistical practitioner(s).
10. Objectively, accurately, and efficiently communicating a team's or practitioners' statistical work throughout the organization.
11. In cases where ethical issues are raised, representing them fairly within the organization's leadership team.
12. Managing resources and organizational strategy to direct teams of statistical practitioners along the most productive lines in light of the ethical standards contained in these Guidelines.